\documentstyle[11pt,bospasp,twoside,epsf]{article} 
\def\edcomment#1{\iffalse\marginpar{\raggedright\sl#1\/}\else\relax\fi}
\reversemarginpar

\begin{document} \title{Dark Matter in Galaxies: Observational overview}
\author{A. Bosma} \affil{Observatoire de Marseille, 2 Place Le Verrier,
13248 Marseille Cedex 4, France}

\begin{abstract} I review the observational side of the present
state of the debate about the dark matter in galaxies, with emphasis on 
the core/cusp problem in low surface brightness galaxies, and the
question of maximum / sub-maximum disks in spiral galaxies. Some 
remarks are made about the dwarf spheroidals around the Milky Way,
and about elliptical galaxies.
\end{abstract}

\section{Introduction}
The WMAP results (Spergel et al. 2003) have given rather strict limits
on the matter and energy content of the Universe, in agreement with the
$\Lambda$CDM model. Typically, 
for $\Omega$$_{\rm total}$ = 1,
$\Omega$$_{\rm dark~energy}$ = 0.73, and $\Omega$$_{\rm matter}$ = 0.27, 
with 
$\Omega$$_{\rm baryon}$ = 0.044, and $\Omega$$_{\rm non-baryonic}$ = 0.23.
Some of these results depend on data from the 2dF galaxy redshift survey.
Initial results from the SDSS redshift survey are not 
much different (Tegmark
et al. 2003). The ratio of baryonic to non-baryonic matter is thus
1 to $\sim$ 5 - 6. The baryonic matter at z = 0 is mostly not in stars 
($\Omega$$_{\rm stars}$ = 0.005), and the matter budget in e.g. the 
Local Group is still not well known : there could be hot gas in 
substantial amounts (Nicastro et al. 2003).

These results further support the $\Lambda$CDM model for structure 
formation in the Universe, and underscore the need to
understand galaxies in the framework of this theory. However, at the
scale of galaxies, the ``predictions'' of the $\Lambda$CDM model depend on
numerical simulations, which, despite their sophistication, suffer 
from inadequate resolution, and may miss some of the physics. 
This meeting aims to see how well specific predictions from 
current cosmological simulations of dark matter halos fare on  
issues such as their central density profiles, 
the relation between dark to baryonic matter there, and their shape and extent.

\section{Some issues concerning mass models of spiral galaxies}

Over the last 30 years observations, in particular rotation curves
derived from HI data, have clearly established that the ``visible'' 
parts of a spiral galaxy, i.e. the stellar bulge and disk, and the 
HI gas layer, cannot account for the observed high rotation speeds in
the outer parts of spirals, if the stellar mass-to-light ratio is 
constant with radius. This discrepancy is attributed to unseen matter,
often taken to be a (spherical) dark halo. 
Several issues complicate this picture~:  \\

\noindent
{\sl Assumptions going into the derivation of the rotation curve}. A
large fraction of galaxies have a warped HI layer in the outer parts, 
which is modelled with a ``tilted ring'' model in which the orientation 
of the rings is gradually changed to allow for the peculiarities of the 
velocity field. Non-axisymmetric disturbances, such as bars or oval 
distortions,
or peculiar velocities due to spiral structure, are in first instance
ignored, and an azimuthally averaged rotation curve derived. Asymmetries,
if small enough, are likewise ignored. All these assumptions may introduce
systematic errors in the derived rotation curves at the 5 - 15\% level.\\

\begin{figure}
\plotfiddle{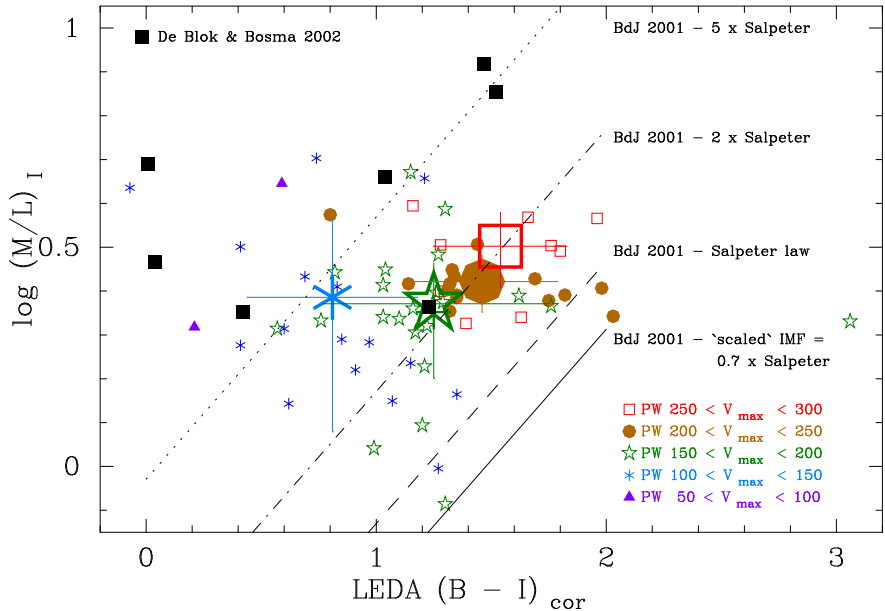}{8cm}{0}{75}{75}{-258}{-323}
\caption{Relation between the mass-to-light ratio in the I-band and
the global (B-I) colour corrected for galactic and internal extinction, 
from the LEDA database, for galaxies in the 
Palunas \& Williams (2000) sample. Different symbols indicate differences 
in the maximum rotation velocity V$_{\rm max}$. Large symbols represent
averages per class of V$_{\rm max}$.  Various lines indicate
models from Bell \& De Jong (2001, table 1), different only in the scaling
of the IMF. Black squares come from data of De Blok \& Bosma (2002) for
LSB galaxies, rescaled to the I-band.} 
\end{figure}

\noindent
{\sl Values of the mass-to-light ratio
are not well determined}. This leads to a degeneracy in the mass models.
At one extreme, the maximum disk approach tries to fit as much of the
mass into the disk as possible, 
without overshooting the rotation curve. For optical
curves, most galaxies can then be fitted without a dark halo,
due to the limited extent of the curve.  In Figure 1  I plot,
for a sample of galaxies studied by Palunas \& Williams (2000),
the M/L$_{\rm I}$-ratio
they determined from maximum disk fits against the extinction corrected 
(B-I) colour I collected from the LEDA database. For large
galaxies a trend exists: on average redder disks have higher M/L ratios,
as expected e.g. in models from Bell \& De Jong (2001). Yet there is 
a fair amount of scatter in each class of galaxies selected according to
their maximum rotation velocity V$_{\rm max}$, in particular for smaller
(bluer) galaxies. The required M/L values
are too high compared to those predicted by
Bell \& De Jong's ``preferred'' Initial
Mass Function (IMF),  the ``scaled IMF'' = 0.7 * Salpeter curve, 
based on maximum disk + halo fits for 
Ursa Major galaxies (Verheijen 1997). The extreme maximum disk models
of the galaxies in the Palunas \& Williams sample  
require a bottom heavier IMF.

This problem is aggravated for low surface brightness (LSB) disk galaxies,
as is shown by the few points taken from models of such galaxies by 
De Blok \& Bosma (2002). This clearly illustrates that the dark matter
problem in these galaxies is much worse, and has lead to the assumption
that even a ``minimum disk'' (M/L = 0) is a reasonable approximation
to the real mass distribution. \\

\noindent
{\sl Scaling the HI can produce a good fit instead of a halo.}
Bosma (1978) noticed that the surface density ratio
of HI gas mass to total mass becomes constant with radius
in the outer parts. Reversely, one can thus scale the
curve for the HI gas with a certain factor to fit the rotation
curve, instead of with a halo. Hoekstra
et al. (2001) did this 
for a number of spirals, and Swaters (1999) for a number
of dwarf galaxies. The scaling factor is not constant,
and varies with the maximum velocity 
of a galaxy (cf. Figure 2).
For a number of galaxies the fits may well break down at
larger radii if more sensitive HI data become available.

\begin{figure}[h]
\plotone{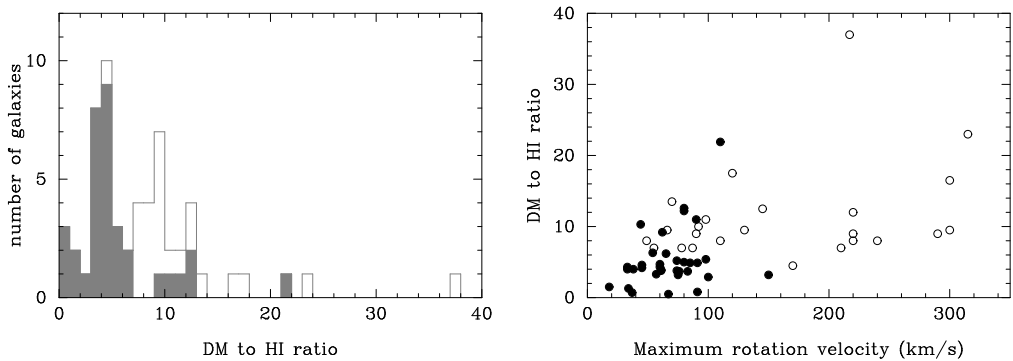}
\caption{HI component scaled as ``dark halo'' : a) histogram
of the scaling factor,
b) scaling factor vs. V$_{\rm max}$. Based
on data from Hoekstra et al. (2001) (blank part of histogram, open circles), 
and Swaters (1999) (filled part of histogram, filled circles)}
\end{figure}

\section{The core/cusp problem in LSB galaxies}

Cosmological numerical simulations invariably produce cuspy dark
halos, but the precise value of the inner slope $\alpha$ of the
radial density profile $\rho \sim {\rm r}^{\alpha}$ is debated. 
Moore et al. (1999) and Fukushige \& Makino (2001)
advocate $\alpha$ = -1.5, and Navarro, Frenk \& White (1996,
1997) an inner slope of -1.0. Both White and
Navarro (this volume) report on work by Hayashi et al. 
(2003) : there is no real convergence towards a unique value of 
the inner slope, the resolution prohibits 
predictions to be made inside a radius 
of $\sim$ 1 kpc, but the profile remains cuspy, and there is 
some cosmic scatter (see also Fukushige, Kawai \& Makino 2003).

This situation differs from an until recently widespread perception  
that the NFW profile, with inner slope -1.0, should be considered 
``universal'', and the yardstick against which the 
observations should be discussed. I will use the slope -1.0 here
as a fiducial mark to see where the observations stand,
and which is to be improved as the cosmological numerical simulations
become more realistic (e.g. by having adequate resolution on the dwarf
galaxy scale, and perhaps incorporating more relevant physics),
and better understood theoretically (e.g. by demonstrating which
physical process, or combination of processes, sets up the 
value of the inner slope in dark matter dominated galaxies).

\subsection{Inner slope values : technical and selection issues}
 
Most observers conclude that $\alpha$
in LSB galaxies is closer to 0.0 than to -1.0, and almost all find
that the decomposition mass models work better if the dark halo
is modelled with a (pseudo-) isothermal sphere, or a Burkert profile,
rather than with the NFW profile. However, some authors maintain
that the problem of determining the inner slope is fraught with
systematic effects, which all magically contribute towards shallower
slopes. The issue debated is not only about the best value for the 
inner slope, but also whether a slope of -1.0 can be 
reasonably excluded, so as to force a modification of the 
$\Lambda$CDM picture at galaxy scales.
See also the contributions by De Blok and by Swaters (this volume).

On the selection side, the debate is about the quality and 
relevance of the data. As 
more galaxies are observed, stricter selection criteria
can be used to retain only galaxies with sufficiently ``good'' data.
Thus one excludes 1) galaxies with poor angular resolution, thereby requiring
in most cases supplementary H$\alpha$ data in addition to a HI
rotation curve, so that the inner 1 kpc of a galaxy is well probed.
This excludes far away galaxies or galaxies for which there is only
low resolution HI data, and thus minimizes slit width effects;
2) edge-on galaxies, which are apparently too difficult to understand
for some workers in the field, despite demonstrations by Bosma
et al. (1992) and Matthews \& Wood (2001) that small galaxies seen
edge-on are transparent and have rotation curves which are slowly
rising in the inner parts. The argument that somehow there is no
emission at the tangent point leads to very peculiar H$\alpha$ 
distributions in more face-on galaxies which are yet to be seen;
3) galaxies with large asymmetries, faint emission, etc., which
have low quality rotation curves.

De Blok, Bosma \& McGaugh (2003) applied these criteria to data of
McGaugh et al. (2001) and De Blok \& Bosma (2002), and find for a
restricted sample that the values of the inner slope are between
0.2 and -1.0 (see De Blok, this volume). 
Swaters et al. (2003a), in a similar study, come to a
similar conclusion. De Blok et al. (2003) also
report about deliberate slit offsets and misalignments
for one galaxy, UGC 4325, and conclude that these effects
could not magically bring the slope from about -0.2 to -1.0, although
of course scatter is introduced if the major axis is not observed
correctly. 

Finally, for several galaxies there are more than one observation, done
with different telescopes, and by different observers. Even though the
error bars are sometimes unsatisfactory large, there is good agreement
in general. The situation for UGC 4325, debated at the meeting, is shown
in Figure 3. The HI position-velocity diagram prompts me to think that
the peak velocity is indeed around $\sim$ 110 km/s as it is in the optical, 
after which the rotation curve declines.

\begin{figure}[t]
\plotfiddle{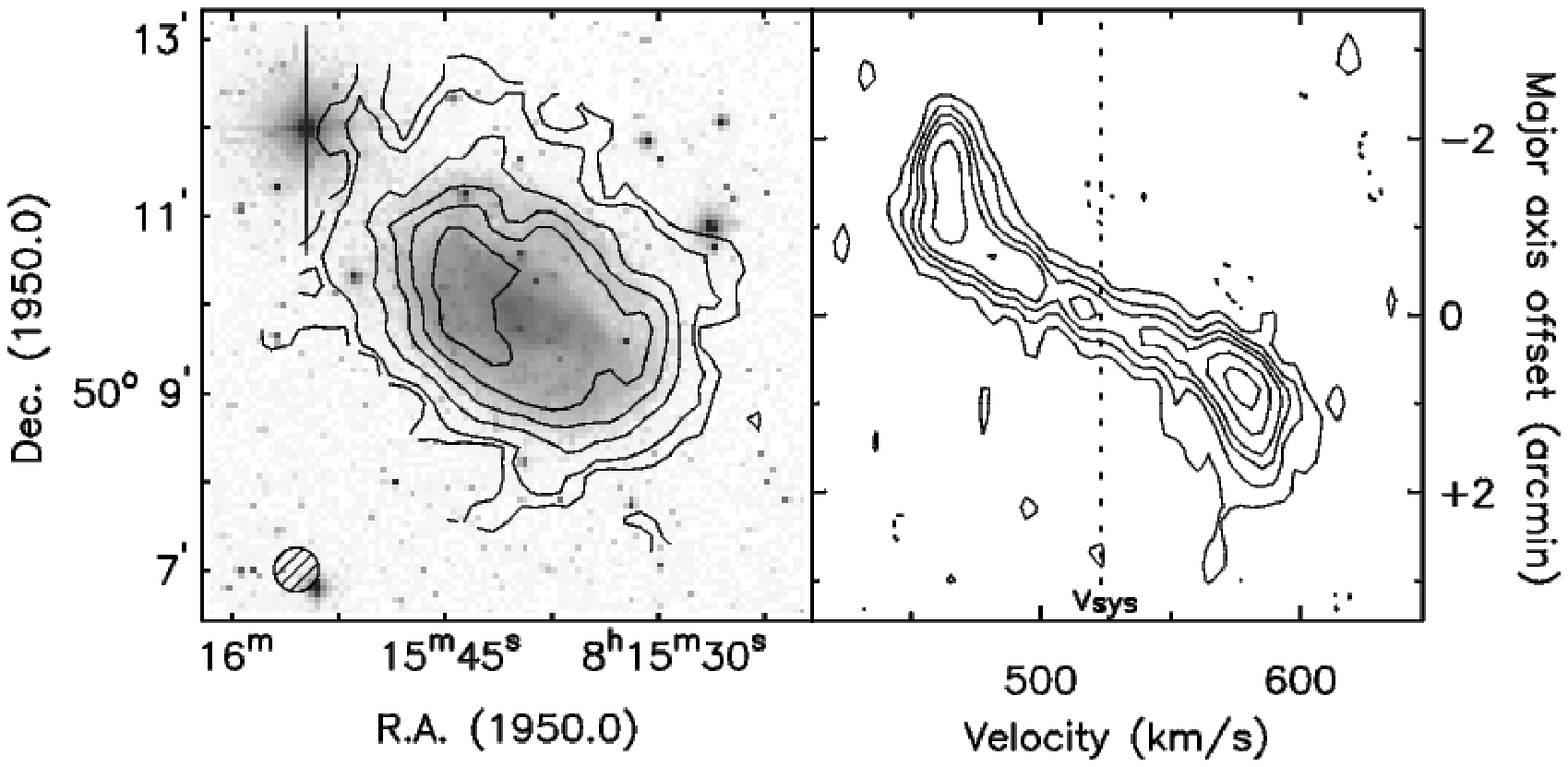}{3.5cm}{0}{35}{35}{7}{-108}
\plotfiddle{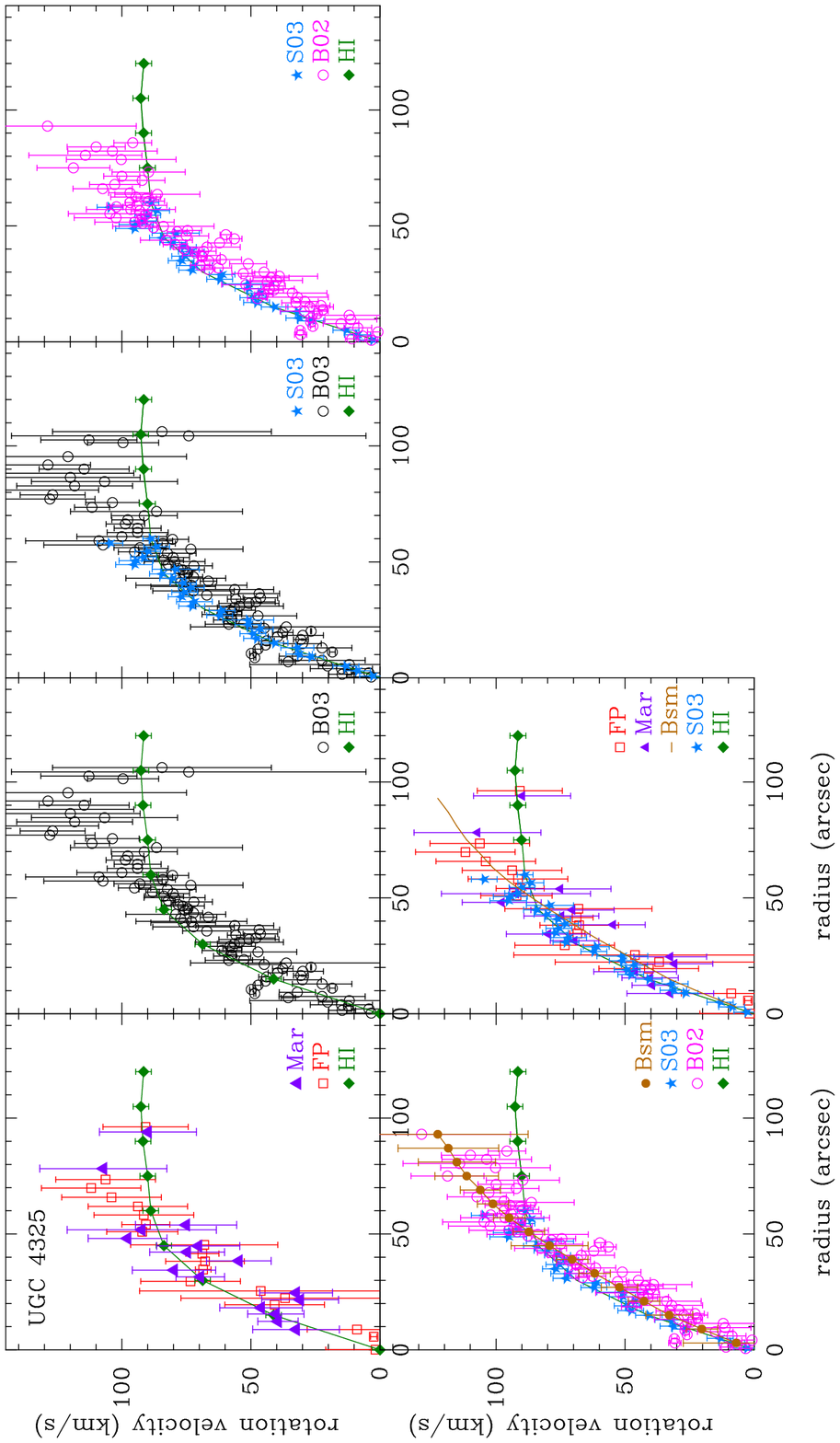}{3.5cm}{-90}{58}{58}{-262}{283}
\caption{Comparison of data for UGC 4325. HI data from Swaters 1999 (HI),
Fabry-P\'erot data from Garrido et al. 2002 (FP), long slit data from 
Marchesini et al. 2002 (Mar), De Blok \& Bosma 2002 (B02), De Blok
et al. 2003 (B03), and Swaters et al. 2003a (S03). Bsm represents
smoothed long slit data as described in B02. 
At lower right the HI distribution and 
position-velocity diagram from Swaters 1999.}
\end{figure}

\subsection{Inner slope values : astrophysical issues}

The determination of the kinematical center, and the influence
of non-circular motions on the rotation curves are also debated.

The rather irregular nature of some of the dwarf or LSB galaxies
studied, and the faintness of particularly the near-infrared images
makes the determination of the center of such galaxies difficult.
Interestingly, HST data on central star clusters in late type spirals is 
available for a number of the galaxies used in the
core/cusp studies, and comparison of these positions and the 
centers of the inner disks show agreement to within 1 - 2 arcsec 
(B\"oker et al. 2002).

For the galaxy IC 2574, I reanalyzed the HI data obtained by Walter \&
Brinks (1999), and deliberately varied the position of the center. In
some cases I also carved up the galaxy into annuli, and allowed the 
center of each annulus to vary with respect to the next one. 
Despite the variation of the center positions with an r.m.s.
scatter of about 250 pc, the determined slopes still come out to be
-0.15 $\pm$ 0.38, well away from the fiducial NFW slope of -1.0.

Bars could contribute to non-circular motions in the inner parts, and
thus lead to a poor quality rotation curve when long slit 
data are used, or when a 
two-dimensional velocity field is not properly analyzed.
Swaters et al. (2003a) assert that bars in their long slit data sample
predominantly have shallow slopes, but inspection shows that UGC
2259, UGC 4499, and UGC 5721, as well as F568-3 are also barred, 
in addition to the
galaxies they identify as such. For the remaining unbarred galaxies,
there is a wide variety of slopes. Swaters et al. (2003b) observed the
galaxy DDO 39 with an integral field spectrograph
yielding a twodimensional velocity field,
which shows that the inner parts of this galaxy are affected by non-circular 
motion caused by a bar-like distortion.
Several other studies of twodimensional H$\alpha$ velocity fields are 
underway. Garrido et al. (2002) report on data obtained with a
Fabry-P\'erot instrument, and Bolatto (this meeting, see Simon
et al. 2003) report on
data obtained with an integral field spectrograph,
combined with CO data obtained with BIMA. The results sofar
for NGC 4605 and NGC 2976 are that the inner slopes are rather shallow,
and that the non-circular
motions are mainly in the inner parts.

One can look at bars in LSB galaxies in two ways: 1) they
could be similar to bars in HSB galaxies, or 2) they could be just the
response to the expected triaxiality of the dark halo. Since the reported
non-circular motions are stronger in the inner parts for DDO 39 and
NGC 2976, it seems likely that those galaxies behave like HSB's, where
the disk is dynamically important in the inner parts. Yet it cannot be
excluded that whole disks are elongated due to the triaxial halo. One way
to study this further is to look at results of a Fourier analysis
of velocity fields of a number of galaxies, performed by Schoenmakers (1999).
From his work, I have collected the results on disk elongation
in Figure 4a, which shows that the scatter in $\epsilon$$_{pot}$ sin $\phi$
is larger for galaxies with smaller maximum rotation velocity. Thus
disks in small galaxies could be globally elongated due to the forcing
of a triaxial halo. Interestingly, the small galaxy with no elongation,
NGC 247, was declared in Van den Bosch et al. (2000) ``the only galaxy 
found [sofar] capable
of limiting the slope of the dark matter profile''. True enough, its
slope is -1.02, close the the NFW slope, but perusal of Schoenmakers's
results and a near infrared image show the presence of a bar, and a 
strong lopsidedness.

In future, the effort will thus shift towards evaluating the importance of
the disc component in LSB galaxies, using near-infrared surface photometry,
and twodimensional velocity fields at high spatial resolution. 
Meanwhile, some evaluation of the importance of non-circular motions can
be had by studying bars in HSB galaxies. Disk elongation
can be constrained using the Tully-Fisher relation (Franx \& de Zeeuw
1992). Early work on
the barred spiral NGC 5383 (Athanassoula 1984) 
shows that the viewing angle is important : some
angles are more favourable than others. Neither the mean rotation curve,
nor the position velocity cut along the major axis can be trusted as a good
representation of the true rotation curve, but the deviations scale with
the strength of the bar. For the weak bars in LSB galaxies, it is not
obvious how strong these effects are.

\section{The importance of disk self gravity in HSB spirals}

This so-called ``maximum disk'' problem is related
to the core/cusp problem, since in the current
$\Lambda$CDM picture the dark halo 
dominates the potential in the central parts also in high surface brightness
spirals. Several ways have been explored to break the mass model
degeneracy using other dynamical considerations 
concerning the importance of the disk in the inner parts of spiral galaxies.

%\subsection{Swing amplifier criteria, amplitude of velocity perturbations}

Athanassoula et al. (1987, ABP) use swing amplifier criteria, 
which depend on the rotation curve shape
and on a characteristic X parameter 
dependent on the epicyclic frequency $\kappa$, the number of arms m, 
and the active surface mass density of the disk.
By requiring that the swing amplification of the m = 2 perturbations
is possible, the range of mass-to-light ratios is limited to 
a factor of 2 : a lower limit set by requiring that the disk is massive 
enough to just allow amplification of the m = 2 perturbations, and an
upper limit set by requiring that amplification of the m = 1 perturbations 
is just prohibited. Usually the latter condition holds for a model 
with maximum disk and a non-hollow halo.

\begin{figure}[t]
\plotone{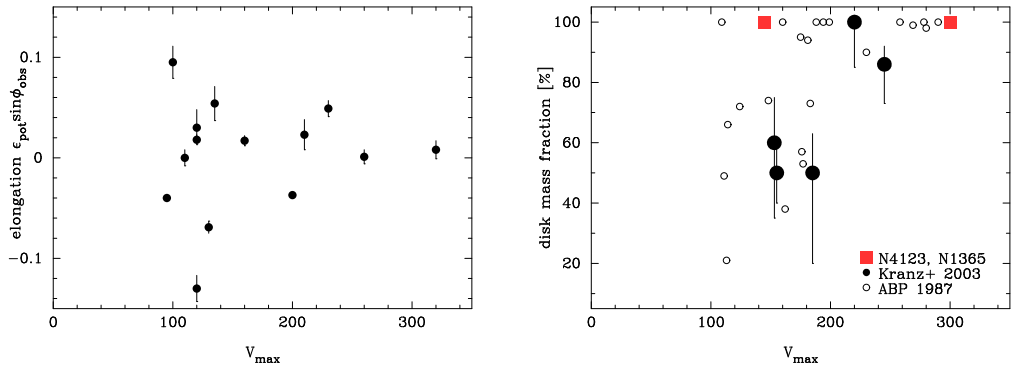}
%\plotfiddle{bosma-fig4.eps}{4cm}{0}{80}{80}{-250}{-472}
\caption{a) Disk elongation as function of V$_{\rm max}$, data extracted 
from Schoenmakers (1999) b) Disk mass
fraction as function of V$_{\rm max}$, data from Kranz et al. (2003) with
N4123 (left) and N1365 added. Faster rotators seem to have rounder disks,
which are more self-gravitating. }
\end{figure}

Peculiar motions due to the spiral
arms were clearly seen in the early M81 21-cm line data,
and modeled with a spiral density wave
response calculation by Visser (1980), who did not include a dark halo 
in his models. The presence of ``wiggles'' in position-velocity curves 
from long slit data are thus associated with the
spiral arms. 
Kranz et al. (2001) use such data for NGC 4254 and try to 
reproduce the observed velocity perturbations with
a stationary gas flow model using the K-band image of this
galaxy as input to the evalution of the disk part of the galactic 
potential. They find that a maximum disk model produces too
large velocity perturbations, and put an upper limit on the
disk mass fraction (the mass ratio between a given disk model and the  
maximum disk model) of 0.8.
However, this galaxy is lopsided in the HI, the spiral may be evolving,
the small bar in the center of the galaxy might have a different
pattern speed than the main spiral pattern,
the inclination may be higher than the authors take it,
and the adopted method may favour lower disk mass fractions (Slyz et al.
2003).
Kranz et al. (2003) report on a similar analysis for four more cases, and 
find a trend that the brightest spirals (those with the highest rotational 
velocities), seem to have maximum disks,
but that towards lower luminosity spirals the relative influence
of the dark matter in the inner parts increases.
Comparison with data from ABP shows good agreement with this trend 
(cf. Figure 4b).

Weiner et al. (2001) model the stationary
gas flow in the barred
spiral NGC 4123, using a potential derived from an optical image,
and find that the best fit to the velocity data
requires a maximum disk model for the mass distribution.
Lindblad, Lindblad \& Athanassoula (1996) find likewise a relatively 
good fit for the bright barred spiral NGC 1365 
with a maximum disk model. Figure 4 suggests that faster rotators 
have rounder disks,
which are more self-gravitating.

A view not necessarily in contradiction
is voiced by Courteau et al. (2003),
who contend that {\it on average}, 
disks with V$_{\rm max}$ $<$ 200 km/s are sub-maximal.
They argue this on the basis of velocity dispersion data from Bottema
(1997) -- which I deem having too large error bars --, work on disk stability 
by Fuchs -- yes, but see Fuchs (2002) --, the absence of the
expected correlated scatter in the Tully-Fisher relation (Courteau \& Rix
1999) -- but disk maximality seems to depend 
on V$_{\rm max}$ -- , and the result on the lens 2237+0305. 
Nevertheless, they find that Tully-Fisher relations for barred and 
un-barred galaxies are similar, in agreement with
previous work, so barredness does not affect maximality. 

Trott \& Webster (2002) combine for 2237+0305 their 
lens model with HI rotation data further out. There is little 
need for a dark halo in the central parts, which are dominated by a 
bulge-bar system. Their statement that the
disk is not maximal is partly influenced by their inclusion of
the bar into the bulge, even though bars are thought to originate
in the disk. For our own Galaxy, data on the microlensing towards
the bulge-bar system likewise suggest that
dark matter does not dominate in the central
parts
(Bissantz \& Gerhard 2002). 

%NGC 4414 fits.

\section{Dwarf spheroidals and other satellites}

The $\Lambda$CDM simulations predict the presence
of far more satellite galaxies than is observed around the Milky Way, 
and the concept of failed dwarfs has been advanced to rescue 
the ``theory''. In this picture, studies of
high velocity clouds are debated, but their individual distances
are quite uncertain, and they don't indeed seem to have stars (e.g.
Simon \& Blitz 2002).

The dwarf spheroidal companions of the Milky Way are very important
in another way : the smallest ones are dark matter dominated.
New data on the Draco system (Kleyna et al. 2002) show that this
galaxy is more extended than previously thought; 
Stoehr et al. (2002) produce a good fit with $\Lambda$CDM. But for
Ursa Minor there is a subpopulation of stars with low velocity dispersion.
Such a cold clump will survive easily in a cored halo potential, but
breaks up rather rapidly in a cusped halo potential (Kleyna et al.
2003).
  
Finally, modelling 
the trail of the Sagitarius dwarf galaxy (e.g. Ibata et al. 2001)
suggests that the dark halo around the Milky Way is nearly spherical.

\section{Elliptical galaxies}

Stellar kinematics of elliptical galaxies
have turned up only 3 unambiguous cases where dark
matter is needed to fit the data : NGC 2434,  NGC 7507 and NGC 7626 
(Kronawitter et al. 2000), partly because the data is limited in
radius to 1 - 2 R$_{\rm e}$. 
Planetary nebulae can now be used as a tracer
further out, and is becoming an industry. Interestingly,
constant M/L models can explain the new data, though M/L values are 
rather high (e.g. Romanovsky
et al. 2003). 

For field ellipticals, evidence for dark matter
does come from HI studies, which
show for several galaxies a flat rotation curve from $\sim$ 0.3 R$_{\rm e}$ 
out to 5 - 6 R$_{\rm e}$
(e.g. Oosterloo et al. 2002). This implies a global
(M/L)$_{\rm B}$ ratio $\sim$ 25. This value is in good apparent 
agreement with a similar value found based on X-ray studies. Yet the new
Chandra and XMM data show a wealth of detail in the images of the X-ray
gas of individual galaxies, so much so that one can question the validity
of the hydrostatic equilibrium equation used to evaluate masses. 
A further complication is the contribution to the
X-ray flux of low mass X-ray binaries, occuring presumably in globular
clusters. Indeed, Paolillo et al. (2003) argue
that the core of the X-ray emission is associated with the stellar
distribution (gas from mass loss of evolving stars), and that the
extended X-ray emission traces really the group or cluster potential
rather than the potential of the elliptical galaxy itself. This agrees
with studies using globular clusters as a tracer, e.g. for NGC 4472 and M87,
which show that there is need for dark matter at
larger radii there (see Kronawitter et al. 2000 and references therein).

A most interesting development is the use of lensing data, combined with
more classical spectroscopy, to estimate the mass of ellipticals. A very
good example of this is the study of the lens 0047-281 by Koopmans \& Treu
(2003), who find indications that 1) the total density follows a power
law with slope -1.9, 2) a constant M/L model can be eliminated using
velocity dispersion data, and 3) the inner dark matter slope can be as
shallow as -1.1. For some lenses, the flux ratios cannot be matched with
a smooth potential, and therefore substructure needs to be added to the
models to produce a fit (e.g. Bradac et al. 2002). This is seen as consistent
with the $\Lambda$CDM prediction.

\acknowledgements{I thank Lia Athanassoula for frequent
discussions, and Erwin de Blok and Stacy McGaugh for fruitful
joint work on LSB galaxies.}

\vfill\eject

\newpage \section*{Appendix: 21-cm Line Interferometry and the Dark
Matter Problem}

\vspace{-0.15truecm} {\it [Reprinted from the Magellanic Times, Day 7,
22 July 2003 - the newspaper edited during the 25th General Assembly of
the IAU in Sydney], minor editing afterwards in {\rm []}.}

\medskip\noindent ``Write me a piece, Albert.  I'd really appreciate it,
I won't have time to do it myself."  I am sitting opposite Seth Shostak,
editor of the Magellanic Times, whom I have known for a long time, and
known of for even longer. He and I owe a lot of intellectual indebtness
and gratitude to a quiet man who patiently taught us the principles of
21-cm HI interferometry, David Rogstad.  Rogstad was one of the first
to use a two element radio aperture synthesis interferometer to study
the 21-cm line emission of nearby galaxies (he was LOC co-chair John
Whiteoak's first student.)

Seth did his thesis on detailed synthesis observations of 5 late type
galaxies with the Owens Valley Radio Observatory (OVRO) twin element
interferometer.  Dave Rogstad moved for a few years to the Kapteyn
Laboratorium in Groningen in the late sixties, where he guided a couple
of students in various projects - Henk Olthof on warps, Arnold Rots on
data of NGC 6946 and IC 342, and me.  He himself worked on his OVRO data
of M101.

For me, it was my first research project.  I had to read reams of charts
to mark off phase zero measurements of fringes from the just completed
Westerbork telescope. Rogstad taught me computing and offered valuable
advice (``beware, if you start computing, you stop thinking.") After this,
he gave me his 21-cm line data of M82 to reduce, along with all his
programs - on punched cards.  I was enthusiastic about this, looked up
the literature on this ``exploding galaxy", and learned about the galaxy
rotation work of Burbidge, Burbidge and Prendergast.  The M82 HI data
were interesting, and Dave encouraged me to speak about them at a YERAC
(Young European Radio Astronomy Conference) in Dwingeloo.

Dave's work on M101 stimulated the young staff members in Groningen, led
by Ron Allen, to push for a temporary receiver to do 21-cm line work on
nearby galaxies with the Westerbork telescope - which was not explicitly
constructed for this purpose.  By the time this project was completed,
I was fortunate enough to use it for my thesis work on a pilot HI study
of spiral galaxies of various morphological types.  That was more or
less [considered]
a backwater project, the main attention for 21-cm line work being
the study of spiral structure of M51, M81 and M101 (``testing the density
wave theory.")

Even so, I witnessed the debate about rotation curves in the outskirts
of galaxies using HI data. I vividly remember a seminar by Mort Roberts,
arguing that M31's rotation curve was flat, and all the young Turks of
the Kapteyn lab arguing that he picked up signal of the main disk through
the sidelobes of his single dish antenna (Ed Salpeter was subjected to
a similar treatment a few years later [at the IAU Symposium 77 in Bad
M\"unstereiffel] when he reported on his Arecibo
data, which had some after-effects if his Annual Reviews autobiography
is any guide).  So 21-cm line interferometry was the creed.

My thesis work, with its basic result that the rotation curves of spiral
galaxies of all morphological types stayed flat (or were even still
rising) beyond the optical image, helped settle the debate of the presence
of dark matter in the outer parts of spirals.  I also was fortunate to be
able to go to Palomar with Piet van der Kruit, to get H$\alpha$  rotation
data in the inner parts, and to obtain plates for surface photometry to
determine the luminosity profile and the mass-to-light ratio as function
of radius.

But the questions remain - how was the dark matter debate conducted,
who said what when and why, and who had the ``correct" (i.e. present day
accepted) picture first?  This is not so easy as some people would have
it.  In Historical Development of Modern Cosmology (ASP conference series
Vol. CS-252), I read two articles - one by Sidney  van den Bergh and one
by Jaan Einasto -  and there seems to be little overlap.  So there could
be different stories, and different people were convinced at different
times of the presence of the dark matter in spirals - just like the
various events and characters which made up the French Revolution. In
any case, as far as the HI line rotation curves is concerned, I can only
cite Vera Rubin et al.'s (1978) paper where she writes that ``Mort Roberts
and his collaborators deserve credit for first calling attention to flat
rotation curves", credit he got only sparingly.  What is clear now is that
there is little need for a dark halo to fit the optical rotation curve
data, a conclusion [again] reached in an ApJ paper as recently as 2000. Mort
Roberts himself confided to me once that he thought that my thesis work
vindicated his contention that rotation curves of giant spirals are flat.
Ostriker and Peebles, who pushed the idea for dark halos, first in a 1973
paper on disk stability (a paper which nowadays cannot withstand close
scrutiny), and then, with Yahil, in their classic 1974 paper on various
indicators pointing to a linear increase of the mass of a giant galaxy
with radius, cite for the rotation data the papers by Roberts and Rots
(1973) and Rogstad and Shostak (1972).

Which brings me back to Seth Shostak. Contrary to my thesis, which I had
printed in 500 copies, and with its handy format became quite in demand
(my thesis advisor had his copy stolen from his office somewhere in
the 1980s, he told me recently), Shostak's thesis (1971) is not widely
available.  I found a copy of it in the VLA library, so I started reading
it carefully, and yes, it contained the flat rotation curve of NGC 2403,
and its implication that the volume density of matter drops off as r$^{-2}$
(mass rises linearly with radius, therefore).  So he had it! Okay,
only for one case, but even so; and it was a clean observation with an
interferometer, so unlike Roberts' observation of M31 (the primary piece
of evidence in the Roberts and Rots (1973) paper).  Oh well, it is unclear
to me sometimes how scientific evidence is accepted in certain circles.
Sometimes one has a result, correct after all, but at a time too far
ahead of the pack, which is not yet ready to absorb it.

In any case, the amusing part of Seth's thesis is in the very end.
A brief acknowledgement to Gordon Stanley, then director of OVRO.
A warm thanks to David Rogstad for his patient advice.  And then this
fantastic flight of fancy - ``this thesis is dedicated to NGC 2403 and its
inhabitants, to whom copies can be furnished at cost".  No wonder this man
is now working on SETI! He just wants to know why nobody came along thus
far to claim a copy.  If somebody from that galaxy had come along, Seth
would have been the most famous man on earth, offered an autographed copy
of his thesis for free, and arranged for a public lecture at the current
General Assembly.  John Whiteoak would have been happy to schedule that,
and the Harbourside Auditorium would have been way too small!

%CAPTION:  Albert Bosma, entirely baryonic.


\begin{references}
\reference Athanassoula, E. 1984, Physics Reports, 114, 319
\reference Athanassoula, E., Bosma, A., \& Papaioannou, S. 1987, \aap, 179, 23
\reference Bell, E., \& de Jong, R.S. 2001, ApJ, 550, 212
\reference Bissantz, N., \& Gerhard, O. 2002, MNRAS, 330, 591
\reference B\"oker, T., et al. 2002, AJ, 123, 1389
\reference Bosma, A. 1978, PhD Thesis, University of Groningen
%\reference Bosma, A. 1981, AJ, 86, 1825
\reference Bosma, A., Byun, Y.I, Freeman, K.C., \& Athanassoula, E. 1992, 
                ApJ, 400, L21
\reference Bottema, R. 1997, \aap, 328, 517
\reference Bradac, M., Schneider, P., Steinmetz, M., Lombardi, M., King, L.J,
                \& Porcas, R., 2002, \aap, 388, 373
\reference Courteau, S., Andersen, D.R., Bershady, M.A., MacArthur, L.A.,
                 \& Rix, H.-W. 2003, ApJ, 594, 208
\reference Courteau, S., \& Rix, H.-W. 1999, ApJ, 513, 561
\reference de Blok, W.J.G., \& Bosma, A. 2002, \aap, 285, 816
%\reference de Blok, W.J.G., McGaugh, S.S., Bosma, A., \& Rubin, V.C. 2001,
%           \apj, 552, L23
\reference de Blok, W.J.G., Bosma, A., \& McGaugh, S.S. 2003, MNRAS, 340, 657
\reference Franx, M., \& de Zeeuw, P.T. 1992, ApJ, 392, L47
\reference Fuchs, B. 2002, astro-ph/0212485
\reference Fukushige, T., \& Makino, J. 2001, \apj, 557, 533
\reference Fukushige, T., Kawai, A., Makino, J. 2003, astro-ph/0306203
\reference Garrido, O., Marcelin, M., Amram, P., \& Boulesteix, J. 2002, 
                \aap, 387, 821
\reference Hayashi, E. et al. 2003, astro-ph/0310576
\reference Hoekstra, H., van Albada, T.S., \& Sancisi, R. 2001, MNRAS, 323, 453
\reference Ibata, R., Lewis, G.F., Irwin, M., Totten, E., \& Quinn, T. 2001
                ApJ, 551, 294
\reference Kleyna, J.T., Wilkinson, M.I., Evans, N.W., Gilmore, G., \& 
                Frayn, C. 2002, MNRAS, 330, 792
\reference Kleyna, J.T., Wilkinson, M.I., Gilmore, G., \& Evans, N.W. 2003,
                ApJ, 588, L21
\reference Koopmans, L.V.E., \& Treu, T. 2003, ApJ, 583, 606
\reference Kranz, T., Slyz, A., \& Rix, H.-W. 2001, ApJ, 562, 164
\reference Kranz, T., Slyz, A., \& Rix, H.-W. 2003, ApJ, 586, 143
\reference Kronawitter, A., Saglia, R.P., Gerhard, O., \& Bender, R. 2000,
                   \aaps, 144, 53
\reference Lindblad, P.A.B., Lindblad, P.O., \& Athanassoula, E. 1996, 
                   \aap, 313, 65 
\reference Marchesini, D. et al. 2002, ApJ, 575, 801
\reference Matthews, L.D., \& Wood, K. 2001, ApJ, 548, 150
\reference Moore, B., Quinn, T., Governato, F., Stadel, J., \& Lake, G. 1999,
                 \mnras, 310, 1147
\reference McGaugh, S.S., Rubin, V.C., \& de Blok, W.J.G. 2001, AJ, 122, 2381
\reference Navarro, J.F., Frenk, C.S., \& White, S.D.M. 1996, ApJ, 462, 563
\reference Navarro, J.F., Frenk, C.S., \& White, S.D.M. 1997, ApJ, 490, 493
\reference Nicastro, F. et al. 2003, Nature, 421, 719
\reference Oosterloo, T.A. Morganti, R., Sadler, E.M., Vergani, D., 
                   \& Caldwell, N. 2002, AJ, 123, 729
\reference Palunas, P.A., \& Williams, T.E.B. 2000, AJ, 120, 2884
\reference Paolillo, M., Fabbiano, G., Peres, G., \& 
                   Kim, D.-W. 2003, ApJ, 586, 850
\reference Romanowsky, A. et al. 2003, Science, 301, 1696
\reference Schoenmakers, R.H.M, 1999, PhD Thesis, University of Groningen
\reference Simon, J.D., \& Blitz, L. 2002, ApJ 574, 726
\reference Simon, J.D., Bolatto, A.D., Leroy, A., \& Blitz, L. 2003, ApJ,
                596, 957
\reference Slyz, A., Kranz, T., Rix, H.-W. 2003, MNRAS (in press), 
                astro-ph/0309597
\reference Spergel, D. et al. 2003, ApJS, 148, 175
\reference Stoehr, F., White, S.D.M., Tormen, G., \& Springel, V. 2002, MNRAS
                335, L84
\reference Swaters, R.A. 1999, PhD Thesis, University of Groningen
\reference Swaters, R.A., Madore, B., van den Bosch, F.C., \& Balcells, M. 
                2003a, ApJ, 583, 732 
\reference Swaters, R.A., Verheijen, M.A.W., Bershady, M.A., \& Andersen, D.R.
                2003b, ApJ, 587, L19
\reference Tegmark, M. et al. 2003, astro-ph/0310273
\reference Trott, C.M., \& Webster, R.A. 2002, MNRAS, 334, 621
\reference van den Bosch, F.C., Robertson, B.E., Dalcanton, J.J, \& de Blok,
                W.J.G. 2000, AJ, 119, 1579 
\reference Verheijen, M.A.W. 1997, PhD Thesis, University of Groningen 
\reference Visser, H.C.D. 1980, \aap, 88, 159
\reference Walter, F., \& Brinks, E. 1999, AJ, 118, 273
\reference Weiner, B., Sellwood, J.A., \& Williams, T.B. 2001, ApJ, 546, 931
\end{references}
\end{document}